\documentclass[11pt]{amsart}
\usepackage{graphicx}
\usepackage{amsfonts, amsmath, amsfonts, amssymb}
\usepackage{algorithm,algpseudocode}
\usepackage{enumerate}
\usepackage{color}
\newtheorem{thm}{Theorem}[section]
\newtheorem{cor}[thm]{Corollary}

\newtheorem{prop}[thm]{Proposition}
\theoremstyle{definition}
\newtheorem{defn}[thm]{Definition}
\theoremstyle{remark}
\newtheorem{rem}[thm]{Remark}
\newtheorem{exmp}[thm]{Example}
\numberwithin{equation}{section}
\numberwithin{equation}{section}

\numberwithin{equation}{section} 

 \numberwithin{equation}{section}
\newenvironment{prf}{ \noindent{\bf Proof}}{\\ \hspace*{\fill}$\Box$ \par  }

\definecolor{BlueClr}{rgb}{0,0,1}
\definecolor{RedClr}{rgb}{1,0,0}

\begin{document}

\title[Geometric Shannon]{Geometric approach to sampling and communication}%
\author{Emil Saucan$^{*}$,  Eli Appleboim$^{\dag}$, 
 and Yehoshua Y. Zeevi$^\dag$}%
\address{Mathematics Department Technion$^*$, Department of Electrical Engineering, Technion$^\dag$}  
\email{semil@tx.technion.ac.il}
\email{eliap@ee.technion.ac.il}%
\email{zeevi@ee.tecnion.ac.il}%

\thanks{First author's research partly supported by the Israel Science Foundation Grant 666/06 and by European Research Council under the European Community's Seventh Framework Programme
(FP7/2007-2013) / ERC grant agreement n${\rm ^o}$ [203134]. }%

\date{\today}%

\maketitle

\begin{abstract}
Relationships that exist between the 
classical, Shannon-type, and geometric-based approaches to sampling are investigated. 
Some aspects of coding and communication through a Gaussian channel
are considered. In particular, a constructive method to
determine the quantizing dimension in Zador's theorem is provided. A
geometric version of Shannon's Second Theorem is introduced.
Applications to Pulse Code Modulation and Vector Quantization of
Images are addressed.
\end{abstract}

\section{General background}

\subsection{Introduction} We consider a geometric approach to
sampling, based on sampling the {\it
graph} of the signal, considered as a {\it manifold}, rather than
sampling in the {\it domain} of the signal. Whereas the latter is widely used in both
theoretical and applied Signal and Image Processing, motivated by
the framework of harmonic analysis, it is important to note that Shannon's original
work is deeply rooted in the geometric approach, at least intuitively \cite{Sh}, \cite{Sh-geo}.
Indeed, this geometric viewpoint of the problem
distinguishes 
Shannon from 
Kotelnikov \cite{Ko} and Nyquist \cite{Ny}, and allows one to
transcend the restricted context of technical communication theory.
We were also inspired in our endeavor by the ``dictionary'' of
geometry-to-communication-theory notions, and we strived to emulate
it. Other paths towards the geometerization of Sampling
Theory can also be found, e.g. in  \cite{Pes} and  \cite{KM}.

Having adopted a geometric approach to sampling, we are concerned with a broad spectrum of signals which can be categorized as {\it geometric signals} -- see Definition 1.3 below. In particular, we are concerned with images that are represented as multidimensional signals and embedded, as such, in $\mathbb{R}^N$, for some large enough $N$. This approach to signals and images has become very popular and well established in recent years, hence it calls for revisiting the basic issues of proper sampling of manifolds and related problems.

Our approach is based on the following sampling existence theorem for
differentiable manifolds that was recently presented and applied in
the context of Image Processing (\cite{SAZ}, \cite{SAZ1})\footnote{A similar
approach to ours appeared in \cite{LL} but,
mathematically less rigorous and comprehensive. Since we
were not aware of this study upon the publication
of our previous works
\cite{SAZ}, \cite{asz}, we use this opportunity to rectify it.}:

\begin{thm} [\cite{SAZ1}]  \label{thm:main-smooth-hd}
Let $\Sigma^n \subset \mathbb{R}^{N}, n \geq 2$ be a connected, not
necessarily compact, smooth manifold, with finitely many compact
boundary components. Then, there exists a sampling scheme of
$\Sigma^n$, with a proper density $\mathcal{D} = \mathcal{D}(p) =
\mathcal{D}\!\left(\frac{1}{k(p)}\right)$, where $k(p) =
\max\{|k_1|,...,|k_{n}|\}$, and where $k_1,...,k_{n}$ are the
principal curvatures of $\Sigma^n$, at the point $p \in \Sigma^n$.
\end{thm}

\hspace*{-0.4cm}(For moore detailed exposition of the Differential Geometry and Topology notions see the Appendix and \cite{doC}).
Here, by ``proper'' density, we mean that it will satisfy the (analogue of) the Nyquist rate, of the classical sampling theory and algorithms.

It is important to note that Theorem \ref{thm:main-smooth-hd} does not necessitate having exact locations of the sampling points.
Thus, the above existence theorem lends itself to a stable constructive algorithm for sampling of manifolds and, therefore, of sampling of higher dimensional signals. Moreover, the following corollary is also applicable to this
problem:

\begin{cor} [\cite{SAZ1}] \label{cor:cor1sampling-hd}
Let  $\Sigma^n, \mathcal{D}$ be as in Theorem \ref{thm:main-smooth-hd}. If there exists $k_0 > 0$,
such that $k(p) \leq k_0$, for all $p \in
\Sigma^n$, then there exists a sampling scheme of $\Sigma^n$ of finite density everywhere. 
%
In particular,
%
if $\Sigma^n$ is compact, then there exists a sampling of $\Sigma^n$
having uniformly bounded density.
\end{cor}

Note, however, that this is not necessarily the optimal scheme (see
\cite{SAZ1}, \cite{asz}).

The constructive proof of this theorem (of which we outline in the Appendix just a sketch) is based on the
existence of the so-called fat triangulations (see \cite{Sa}). The
density of the vertices of the triangulation (i.e. of the sampling)
is given by the 
maximal principal curvature.
 An essential step in the construction of the said triangulations consists of
 isometrically
 embedding of $\Sigma^n$ in some $\mathbb{R}^{N}$, for large enough $N$ (see \cite{Pe}), where the existence of such an embedding
 is guaranteed by Nash's Theorem (\cite{na}). Resorting to such a powerful tool as Nash's Embedding Theorem appears to be an impediment of
 our method, since the provided embedding dimension $N$ is excessively high (even after further refinements due to
 Gromov \cite{gr} and G\"{u}nther \cite{gu}). Furthermore, even finding the precise embedding dimension (lower than the canonical $N$) is very difficult even
 for simple manifolds. However, as we shall indicate in the next section, this high embedding dimension actually
 becomes an advantage, at least from information theoretic viewpoint.

The resultant sampling scheme is in accord with the classical
Shannon theorem, at least for the large class of (bandlimited)
signals that also satisfy the condition of being $\mathcal{C}^2$
curves. In our proposed geometric approach, the radius of curvature
substitutes for the condition of the Nyquist rate. To be more
precise, our approach parallels, in a geometric setting, the {\it
local bandwidth} of \cite{Ho} and \cite{zs}. In other words,
manifolds with bounded curvature represent a generalization of the
{\it locally band limited signals} considered in those papers.
However, the 1-dimensional case is a limiting, degenerated case,
from the geometric viewpoint. As the notion of {\it
fatness} (of simplices), essential to the geometric sampling scheme
(see \cite{SAZ1}), reduces, accounts for dihedral angles, and since on the real line we cannot impose such angles,
the $1$-dimensional version of Theorem \ref{thm:main-smooth-hd} practically reduces to uniform sampling according to the Nyquist rate.

It should be stressed that, in
comparing the classical and geometric approaches, one should bear
in mind that no algebraic structure is presumed in the geometric
context, whereas it is implicitly assumed in the (infinite) sum
appearing in the classical version. Also, it should be noted that, in fact, Shannon
already had the intuition of the role of curvature (using second
partial derivatives) for sampling and begun to explore its geometry
in \cite{Sh-geo}.

Here we further investigate the extent and
implications of this analogy, and of the geometric approach in
general.  We begin by making, in the next Section, a few observations regarding the extent
of our results, by finding the largest space of signals wherein our
results may be applied effectively. Next, in Section 3, we establish
the proper analogies considering concepts originating from classical
sampling
 and from coding theory, considered in the context of Gaussian  channels. In doing so, we attempt to construct a ``dictionary'' of geometric sampling and concepts originating from Shannon's fundamental approach \cite{Sh48}.

As already mentioned, the paper concludes with an Appendix
containing, for the sake of completeness and as a convenient
reference, a brief review of the proof of Theorem 1.1.


\subsection{General geometric signals}

First, let us establish the following definition:

\begin{defn}
A {\it geometric signal} is the graph ${\rm Graph}(f)$ of a function $f:\mathbb{R}^m \rightarrow \mathbb{R}^n$, appertaining to a desired class (e.g. $L^2, \mathcal{C}^2, \mathcal{C}^\infty$), endowed with a (natural) geometric structure.
\end{defn}

The geometric structure, usually considered in vision and image processing, is the that of a Riemannian manifold. This is also the approach we followed in proving Theorem 1.1, as emphasized in the previous section. However, the much larger class of Alexandrov spaces (see, e.g. \cite{BBI}) can be considered in applications (see \cite{asz2}).

\begin{exmp}
A basic example of a geometric signal is that of gray scale images, i.e. surfaces $S$ in $\mathbb{R}^3$, $S = \left((x,y),h((x,y))\right)$, where $(x,y)$ represent the pixel coordinates and the function $h$ represents the gray scale level (intensity). In this case, the usual geometric structure considered is the Riemannian one induced by the ambient $3$-dimensional Eclidean space, at least when $S$ is considered to be a smooth surface. However, if a Combinatorial (discrete) Image Processing viewpoint is adopted, then one can consider $S$ to be endowed only with its metric structure (inherited, again, from $\mathbb{R}^3$), see for instance, \cite{SA}, \cite{SAWZ}.
\end{exmp}

As a followup of the definition and example above, unless stated otherwise, we consider signals as graphs of functions endowed with their natural geometric structure.

We begin our investigation by noting that, by the Paley-Wiener
Theorem (see, e.g., \cite{ra}), any bandlimited signal is of class
$\mathcal{C}^\infty$. We have already shown in \cite{SAZ} that our
geometric sampling method applies not only to bandlimited signals,
but also to more general $L^2$
functions whose graphs are smooth $\mathcal{C}^2$ curves, not
necessarily
planar.
In fact, the geometric sampling approach can be extended to a far
larger class of manifolds. Indeed, every piecewise linear ($PL$)
manifold of dimension $n \leq 4$ admits a (unique, for $n \leq 3$)
{\it smoothing} (see for example \cite{th}), and every topological
manifold of dimension $n \leq 3$ admits a $PL$ structure (cf., for
example, \cite{th}). In particular, for curves and surfaces, one can
first consider a smoothing of class $\geq \mathcal{C}^2$ (so that
curvature can be defined properly), which can then be sampled with
sampling rate defined by the maximal curvature radius. Since a given
manifold and its smoothing are arbitrarily close as smooth submanifolds of some $\mathbb{R}^n$ (i.e. in the {\it fine $\mathcal{C}^1$ topology} -- see  \cite{mun}), one
obtains the desired sampling result. (This very scheme is developed
and applied in \cite{SAZ} for gray-scale images.)

While numerical schemes for practical implementation of smoothing
exist, they are not necessarily computationally satisfactory. For
practical applications, one can circumvent this problem  by applying
numerical schemes based on the finite element method.
However, for the sake of mathematical correctness and in order to be
able to tackle more general applications, one should consider
more general curvature measures (see, e.g. \cite{Z}) and avoid
smoothing altogether (see 
Section 2 below, for a brief discussion of this topic
in a slightly different context).

It is worthwhile to highlight yet another aspect of our geometric
sampling method: Shannon's Sampling theorem relates to bandlimited
signals, that are, necessarily, unbounded in time or in spatial domain.
Obviously, unbounded signals are not encountered in Signal/Image
Processing, nor in any other practical implementations. 
However, geometric sampling does not necessitate having unbounded signals, 
quite the
opposite: it is by far much easier to apply geometric sampling in cases of bounded manifolds.
In practical implementations, this drastically reduces {\em aliasing} effects, while eliminating the need to produce periodic signals
(surfaces).

\section{Important Implications}

\subsection{Pulse Code Modulation for Signals and Images}

Our geometrical approach to sampling lends itself to consideration
of a broader range of topics in communications. In particular,
it offers 
a new method for PCM ({\it pulse code modulation}) of
images considered as high dimensional objects and not as a $1$-dimensional signal obtained by some projection map (scan), of the image to a single dimension. The geometric approach is endowed with an inherent advantage in
that the sampling points are associated with relevant geometric
features (via curvature) and are not imposed arbitrarily at equal intervals determined by the Nyquist
rate. To be more precise, each code word is represented, according to the proposed geometric approach, by a (sampling) point in some $\mathcal{R}^N$, belonging to the given geometric signal. The reconstruction of the signal is, basically, the piecewise flat one described in the Appendix. Of course, better results can be obtained if smoothing of the manifold is considered, especially if this is done using a curvature-based kernel \cite{asz2}. Since this process of smoothing requires one to encode also the curvature of the reconstructed piecewise linear manifold, there should be developed a way to quantize curvature so that the number of extra bits added should be bounded.

Moreover, it should be stressed again that the geometric approach, based on curvature radii, inherently produces a sparse, with respect to the Nyquist rate, adaptive sampling (see \cite{asz}), lending itself to interesting benefits, insofar as various applications are concerned. Indeed, the sampling density is lower in (almost) flat regions. Therefore, fewer sampling points are needed for such regions. (See \cite{asz} for more details.) 
A version of Landau's Theorem \cite{lan2}, \cite{lan3} about
non-uniform sampling is also attained and will be referred to again in Section $3$. Moreover, given that the
theoretical {\it tubular radius} (defined in Section $3$), which is in general practically
impossible to compute, is replaced by the curvature radii, and
since for certain types of manifolds better global estimates may be
obtained in terms of other curvature measures (e.g the {\it Ricci
curvature} (see \cite{asz}), one can in some special cases, lower rates than the {\it principal curvature} rate given in Theorem \ref{thm:main-smooth-hd}.


\subsection{Vector Quantization for Images}

A complementary byproduct of the constructive proof of Theorem
\ref{thm:main-smooth-hd} is a precise method of {\em vector
quantization} or {\em block coding}. Indeed, the proof of Theorem
\ref{thm:main-smooth-hd} consists of construction of a Voronoi
(Dirichlet) cell complex $\{\bar{\gamma}_k^n\}$, whose vertices
provide the sampling points. The set of centers $\{a_k\}$ of the cells,
satisfying a certain geometric density condition, represent, as
usual, the {\em decision vectors}. An advantage of this approach,
besides its simplicity, is the possibility to estimate
the error in terms of length and angle distortion, when reverting from
the cell complex $\{\bar{\gamma}_k^n\}$ to the Euclidean cell
complex $\{\bar{c}_k^n\}$ having the same set of vertices
$\{\bar{\gamma}_k^n\}$ (see \cite{Pe}).
%
%
\subsubsection{Error bounds}
Since the geometric method represented herein enables one to perfect reconstruction of the sampled manifold by an iterative process in which, at each iteration, a piecewise linear approximation is obtained it is important to have error bounds on the approximation at each iteration. It is shown in \cite{SAZ} (see appendix Theorem \ref{s:Munkres}) that this error in the can be bounded in terms of the fatness coefficient of the obtained triangulation (defined in the appendix, Definition \ref{defn:FantTriang}) and of the diameter of simplices.

Yet, for a more accurate error assessment, it is desirable to have bounds on the local distortion of the metric
between the
piecewise-flat approximation and the intrinsic metric of the manifold. Bounding this distortion of metrics bounds the distortion of geometries of the two objects (the approximation and the manifolds) say, in terms of curvature measures, volume diameter and other geometric attributes. This will allow us to say that the reconstructed approximation not only approximate the manifold pointwise, but also gives a good approximation to the geometry of the samples manifold. This is given by the following Theorem, \cite{Pe}
\begin{thm}[\cite{Pe}]
If $M = M^n$ is a manifold without
boundary, then locally, for any triangulation patch, the following
inequality holds,
\begin{equation}
\frac{3}{4}d_M(x,y) \leq d_{eucl}(\bar{x},\bar{y}) \leq  \frac{5}{3}d_M(x,y)\,;
\end{equation}
where $d_{eucl}, d_M$ denote the Euclidean and intrinsic metric (on
$M$) respectively,  and where $x,y \in M$ and $\bar{x},\bar{y}$ are
their preimages on the piecewise-flat complex.
\end{thm}
(The building of these patches is essential for the control of the
fatness of the triangulation. Their size essentially depends upon
the (local) maximal curvature -- see \cite{Pe} and the Appendix.)

For manifolds with boundary we have,
\begin{thm}[\cite{Sa}]
\begin{equation}
\frac{3}{4}d_M(x,y) - f(\theta)\eta_\partial \leq d_{eucl}(\bar{x},\bar{y}) \leq  \frac{5}{3}d_M(x,y)  + f(\theta)\eta_{\partial M}\,;
\end{equation}
where $f(\theta)$ is a constant depending on the $\theta =
\min{\{\theta_{\partial M},\theta_{{\rm{int}}\, M}\}}$ -- the fatness
of the triangulation of $\partial M$ and ${\rm{int}\, M}$,
respectively, and $\eta_\partial$ denotes the {\it
mesh} of the triangulation (i.e. the supremum of the diameters of the simplices
belonging to the triangulation) of a certain
neighbourhood of $\partial M$ (see \cite{Sa}).
\end{thm}

In other words, the (local) projection  mapping, $\pi$, between the triangulated manifold
$M$ and its piecewise-flat approximation $\Sigma$ is (locally) {\it
bi-lipschitz} if $M$ is open, but only a {\it quasi-isometry} (or
{\it coarsely bi-lipschitz}) if the boundary of $M$ is not
empty. (In fact, as the two inequalities above show, the projection mapping $\pi$
 satisfies, in both cases, slightly stronger conditions.) Note that inequalities (2.1.) and (2.2) imply that, both for open and bordered manifolds, control of the Euclidean error of the sampled points is equivalent to the control of the error for the sampled geometric signal.


\subsection{Zador's Theorem}

Yet, a more important benefit of the proposed approach stems from  Zador's Theorem
\cite{Za}. The latter basically states that it is more efficient to use high-dimensional quantizers \cite{CS}.
This result  implies that we can turn into an advantage the inherent
curse of dimensionality. Indeed, by Zador's Theorem, the {\it
average mean-squared-error per dimension} is:

\begin{equation}
\mathcal{E} = \frac{1}{N}\int_{\mathbb{R}^N}{d_{eucl}(x,p)}p(x)dx\,,
\end{equation}
$p_i$ being the {\it code-point} closest to $x$, and $p(x)$ denotes
the {\it pdf} of $x$, can be obtained more efficiently by means of higher dimensional quantizers (see \cite{CS}).

Since for embedded manifolds, it obviously holds that $p(x) = p_1(x)\chi_M$,
where $\chi_M$ is the characteristic function of $M$, defined on the ambient space $\mathbb{R}^N$,
we obtain:

\begin{equation}
\mathcal{E} = \frac{1}{N}\int_{M^n}{d_{eucl}(x,p_i)}p_1(x)dx\,.
\end{equation}
It follows that if the main issue is accuracy, and not simplicity, then
1-dimensional coding algorithms perform far worse than higher dimensional ones. Of
course, there exists an upper limit for the coding
dimension. Otherwise, one could just encode the whole data as
one $N$-dimensional vector for $N$ large enough, which in practice may be huge.
The geometric coding method proposed here provides a
{\em natural} high dimension for the quantization of $M^n$ which is, the
embedding dimension $N$. Moreover, it closes (at least for images
and any other data that can be represented as a Riemannian manifold)
the open problem (related to Zador's Theorem) of finding a constructive
method to determine the dimension of a quantizer (Zador's proof
is nonconstructive). In fact, for a uniformly distributed input (such as a
manifold, and for example, as a noiseless image assumed to be), a better estimate of the average mean-squared-error per dimension can be obtained:

\begin{equation}
\mathcal{E} = \frac{\frac{1}{N}\int_{M^n}{d_{eucl}(x,p_i)}dx}{\int_{M^n}dx} = \frac{\frac{1}{N}\int_{M^n}{d_{eucl}(x,p_i)}dx}{\mathcal{V}_n(M^n)dx}\,,
\end{equation}
where $\mathcal{V}_n$ denotes the $n$-dimensional volume (area) of
$M$. Whence, for compact manifolds, one obtains the following
expression for $\mathcal{E}$:
\begin{equation}
\mathcal{E} = \frac{\frac{1}{N}\int_{M^n}{d_{eucl}(x,p_i)}dx}{\sum_i^m\int_{V_i}dx} = \frac{\frac{1}{N}\int_{M^n}{d_{eucl}(x,p_i)}dx}{\sum_i^m{\mathcal{V}_n(V_i)}dx}\,,
\end{equation}
where $\{V_i\}$ represent the Voronoi cells  of the partition. Moreover, we have the following estimate for the {\it quantizer}: Choose centers of cells such that the quantity

\begin{equation}
\mathcal{Q} = \frac{1}{N}\frac{\frac{1}{m}\int_{M^n}{d_{eucl}(x,p_i)}dx}{\left(\frac{1}{m}\sum_i^m\mathcal{V}_n\right)^{1+\frac{2}{N}}} \;
\end{equation}
is minimized. %

The embedding dimension $N$ increases
dramatically, even for compact manifolds and even taking into
consideration Gromov's and G\"{u}nther's improvements of Nash's
original method (see \cite{gr}, resp. \cite{gu}). For instance, $n =
2$ requires embedding dimension $N = 10$ and $n = 3$ the
necessitates $N=14$. Hence, for large enough $n$ one can write following rough estimate:

\begin{equation}
\mathcal{Q} \approx \frac{1}{N}\frac{\int_{M^n}{d_{eucl}(x,p_i)}dx}{\sum_i^m\mathcal{V}_n}\,.
\end{equation}


\section{Sampling and Codes}

\subsection{Packings, Coverings and Lattice codes}
According to classical signal processing theory, the required signal bandwidth $W$
and the {\it Nyquist sampling rate} are constrained by the condition
$W = \eta/2$, where $W$ and $\eta$ are the bandwidth and sampling rate respectively. This lends itself to an immediate generalization to periodic signals, or, in
geometric terms, for signals represented over a {\it lattice}:
$\Lambda = \{\lambda_i\}$. One can even interpret the boundaries of the lattice as the coordinates in a
multi-dimensional (warped) space or time (see, e.g. \cite{S},
\cite{L}). (Alternatively, one can interpret the dimensions
as representing wave-length, or even as combined fundamental
quantities, e.g. space-time or even space-time-wave-length, as they
arise in Medical Imaging (CT).) Note that such signals can be viewed
as distributions on the $n$-dimensional torus $\mathbb{T}^n =
\mathbb{R}^n/\mathbb{Z}^n$. %
According to this interpretation, the ($n$-dimensional!) period is
the {\it fundamental cell} $\lambda$ of the lattice. Two scalars are
naturally associated with this cell: its diameter, ${\rm
diam}(\lambda)$ (or, alternatively, the length of the
longest edge), and its volume ${\rm Vol}(\lambda)$. Either one may be
used as a measure of the $n$-dimensional period. However, they are
both interrelated and associated with one geometric feature, the
so-called ``{\it fatness}'':

\begin{defn}
Let $\gamma = \gamma^k$ be a $k$-dimensional cell. The {\it
fatness} (or {\it aspect-ratio}) of $\gamma$ is defined as:

\begin{equation}
\varphi(\gamma) = \min_{\lambda}{\frac{{\rm Vol}(\lambda)}{{\rm diam}^l(\lambda)}}\,,
\end{equation}
where the minimum is taken over all the $l$-dimensional faces of
$\gamma$, $0 \leq k$. (If ${\rm dim}\,\lambda = 0$, then ${\rm
Vol}(\lambda) = 1$, by convention.)
\end{defn}

In the case of simplices (and, since any cell can be canonically decomposed into
simplices, of regular cells) this definition of
fatness is equivalent to the following one 
(see
\cite{Pe}):

\begin{defn} \label{def:fat-triang}A $k$-dimensional cell $\gamma$ 
is called $\varphi${\em -fat} if there exists $\varphi > 0$
such that the ratio $\frac{r}{R} \geq \varphi$; where $r$ denotes
the radius of the inscribed sphere of $\gamma$ (or {\it in-radius})
 and $R$ denotes the radius of the circumscribed sphere of $\gamma$ (or {\it circum-radius}). 
A cell-complex $\Gamma = \{ \gamma_i \}_{i\in \bf I }$ is {\em fat}
if there exists $\varphi \geq 0$ such that all its cells are
$\varphi$-{\it fat}.
\end{defn}

Recall that the in- and circum-radius are important in lattice
problems: given a lattice $\Lambda$ with (dual) Voronoi cell
$\Pi$ (of volume 1), one has to minimize the in-radius, to solve the packing problem, and to minimize the circum-radius for solving the covering problem (see \cite{CS}). %
Note that $\Lambda$ and $\Pi$ are simultaneously fat.
It follows that fat cell-complexes and, in particular, fat triangulations, 
represent a mini-max optimization for both the packing and the covering problem. 
 Moreover, since fat triangulations are essential for the sampling theorem for manifolds, it appears that there exists an
intrinsic relation between the sampling problem for manifolds and
the covering and packing problems.
%


\subsection{Average Power, Rate of Code and Channel Capacity}
It is natural to extend the classical definitions of {\it average
power} of a signal:

\begin{equation}
P = \frac{1}{T}\int_{0}^{T}f^2(t)dt\,,
\end{equation}
and the {\it rate} of a code:

\begin{equation}
R = \frac{1}{T}\log_{2}{N}\,,
\end{equation}
in the context of lattices with fundamental cell $\lambda$, where
$N$ represents the number of code points, in the following manner:

\begin{equation}
P = \frac{1}{{\rm Vol}(\Lambda)}\int_{\lambda}f^2(t)dt = \frac{1}{N_1{\rm Vol}(\lambda)}\int_{\lambda}f^2(t)dt\,,
\end{equation}
and

\begin{equation}
R = \frac{1}{{\rm Vol}(\Lambda)}\log_{2}{N} = \frac{1}{N_1{\rm Vol}(\lambda)}\log_{2}{N},
\end{equation}
respectively, $N_1$ being the number of cells.

Similarly, one can adapt the classical definition of the {\it
channel capacity}:

\begin{equation}
C = \lim_{T \rightarrow \infty}R = \lim_{T \rightarrow \infty}{\frac{\log_2{N}}{T}}\,,
\end{equation}
to become
\begin{equation}
C = \lim_{T \rightarrow \infty}{\frac{\log_2{N}}{{\rm Vol}(\Lambda)}} = \lim_{T \rightarrow \infty}\frac{1}{N_1{\rm Vol}(\lambda)}\log_{2}{N}.
\end{equation}
Since $N$ and $N_1$ are related by $N_1 = \alpha(N)$, where $\alpha$
is the {\it growth function} of the manifold, the expression of channel capacity, $C$,
becomes:

\begin{equation}
C = \lim_{T \rightarrow \infty}\frac{1}{{\rm Vol}(\lambda)}\frac{\log_2{N}}{\alpha(N)}.
\end{equation}

It follows that $C = \infty$ for non-compact Euclidean and
Hyperbolic manifolds, and $C = 0$ for their Elliptic counterparts.
Unfortunately, no such immediate estimates can be easily produced
for manifolds of variable curvature.

Note that by substituting $1/T = {\rm Vol}(M)$, the above definitions apply to any sampling scheme of any manifold of finite volume, not
just to lattices. In this case $N$ and $N_1$ represent the number
of vertices, respective simplices, of the triangulation. In
the context of classical signal processing, this approach is known
as ``{\it recurrent nonuniform sampling''} \cite{EO}. (See also
\cite{Wa}.)

The interpretation of frequency considered above does not extend,
however, to general geometric signals. For a proper generalization
we have to look into the geometric analogue of $W$. Based on
\cite{SAZ1}, according to Theorem 5.2, for the case of curves, i.e
$1$-dimensional (geometric) signals, $W$ equals the {\it curvature
rate} $k/2$, were $k$ represents the maximal absolute curvature of
the curve. This, and the sampling Theorem 4.11 of \cite{SAZ1},
naturally lend themselves to the following definition of $W$ for general
geometric signals:
\begin{defn}
Let $M = M^n$ be an $n$-dimensional manifold  $n \geq 2$. $W = W_M =
1/k_M$, where $k_M = \max{k_i}$ and $k_i, i=1,\ldots, n$ are the
principal curvatures of $M$.
\end{defn}
According to classical considerations, the energy $E$ of the signal
$f:\mathbb{R} \rightarrow \mathbb{R}$ is considered to be equal to
its $L^2$ norm:

\begin{equation}
E = E(f) = \int_{-\infty}^{\infty}f^2(t)dt = \frac{1}{2W}\sum_{k = -\infty}^{+\infty}f^2\big(\frac{k}{2W}\big)\,.
\end{equation}

One would like, of course, to find proper generalizations of the
notion of energy for more general (geometric) signals. In view of
the above discussion, the first step is to replace
$2W$ by the proper generalized expression. However, when considering more
general function spaces of specific relevance (s.a. {\it bounded
variation} ({\it BV}), {\it bounded oscillation} ({\it BO}), {\it
bounded mean oscillation} ({\it BMO})),
one should consider energies fitting 
the specific norm of the space under consideration. This discussion
is, of course, also valid with regard
to the average power $P$, and rate $R$, of a
geometric signal.

We proceed to consider the first definition of code efficiency: the
{\it nominal coding gain} (ncg) of a code $c_1$ over another code, say
($c_2$), is defined as:

\begin{equation}
{\rm ncg}(C1,C2) = 10\log_{10}{\left(\frac{\mu_1}{E_1}\Big/\frac{\mu_2}{E_2}\right)}\,,
\end{equation}
where $\mu$ is the square of the minimal squared-distance between
coding points. For geometric codes of bounded curvature (hence
compact codes), the expression for $\mu$ is in particular simple: $\mu
= 1/\min{k}$ ($k$ denoting again principal curvature).



\subsection{The Channel Coding Problem}

It is most natural to approach problems associated with the {\it
Gaussian white noise channel} in the context of ``geometric
signals'', i.e. in the context of manifolds. Recall that in the
classical context, a {\it received signal} is represented by a
vector $X = F + Y$, where $F = (f_1,\ldots ,f_N)$ is the {\it
transmitted signal}, and $Y = (y_1,\ldots ,y_n)$ represents the
noise, whose components $y_i$ are independent Gaussian random
variables, of {\it mean} $0$ and {\it average power} $\sigma^2$. The
main, classical result for the Gaussian channel is the following:

\begin{thm} [Shannon's Second Theorem, \cite{Sh}]
For any rate $R$ not exceeding the channel {\em capacity} $C_0$,

\begin{equation}
C_0 = \frac{1}{T}\log_2\left(1 + \frac{P}{\sigma^2}\right),
\end{equation}
there exists a sufficiently large $T$, such that there exists a code
of rate $R$ and average power $\leq P$, and such that the
probability of a decoding error is arbitrarily small. Conversely, it
is not possible to obtain arbitrarily small errors for rates $R >
C_0$.
\end{thm}


In the case of geometric signals, $F$ is given by the sampling
(code) points on the manifolds and, since the mean equals $0$, the
noisy transmitted signal $F + Y$ lies in the {\it tube} ${\rm
Tub}_\sigma(M)$. Recall that tubes are defined as follows:

\begin{defn}

Let $M \subset \mathbb{R}^m$ be an orientable embedded manifold, and let
$\bar{N}_p$ denote the unit normal to $M$ at the point $p$. For each
$p \in M$ consider the open
symmetric interval of length $2\varepsilon_p, \; I_{p,\varepsilon_p}$, in the direction of $\bar{N}_p$,
where $\varepsilon_p$ is to be chosen small enough such that
$I_{p,\varepsilon_p} \cap I_{q,\varepsilon_q} = \emptyset, \; \textrm{\rm for all}\;
p,q \in M, || p - q ||_{2} > \xi \in \mathbb{R}_+$. Then $Tub_\varepsilon(M) =
\bigcup_{\scriptsize p \in M}I_p$ is an open set that contains $M$,
such that for any point $x \in Tub_\varepsilon(M)$ there exists a
unique normal line to $M$ through $x$. $Tub_\varepsilon(M)$ is called
a {\it tubular neigbourhood} of $M$ or just a {\it tube}.
\end{defn}

The existence of tubular
neighborhoods is assured both locally, for any regular, orientable
manifold, 
and globally for regular, compact, orientable manifold (see \cite{G}). 
In addition, the regularity of the  manifolds $\partial{\rm
Tub}_\sigma^-(M) = M - \bigcup_{p \in M}\varepsilon\bar{N}_p$,
$\partial{\rm Tub}_\sigma^+(M)= M + \bigcup_{p \in
M}\varepsilon\bar{N}_p$, where $\partial{\rm Tub}_\sigma^-(M) \cup
\,\partial{\rm Tub}_\sigma^+ (M) = \partial {\rm Tub}_\sigma(M)$, is
at least as high as that of $M$: If $M$ is convex, then
$\partial{\rm Tub}_\sigma^-(M)$, $\partial{\rm Tub}_\sigma^+(M)$ are
piecewise $\mathcal{C}^{1,1}$ manifolds (i.e. they admit
parameterizations with continuous and bounded derivatives), for all
$\sigma > 0$. Also, if $M$ is a smooth enough manifold with a
boundary, that is, at least piecewise $\mathcal{C}^{2}$, then
$\partial{\rm Tub}_\sigma^-(M)$, $\partial{\rm Tub}_\sigma^+(M)$ are
piecewise $\mathcal{C}^{2}$ manifolds, for all small enough $\sigma$
(see \cite{F}).

In the geometric setting, $\sigma$ can be taken, of course, to be
the maximal Euclidean deviation. However, a better deviation measure
is, at least for compact manifolds, the Haussdorf Distance (between
$M$ and $\partial{\rm Tub}_\sigma^-(M)$, $\partial{\rm
Tub}_{\sigma}^+(M)$):

\begin{defn}
Let $(X,d)$ be a metric space and let $A,B \subseteq (X,d)$. The
{\em Hausdorff distance} between $A$ and $B$ is defined as:

\begin{equation}
d_H(A,B) = \max\{\sup_{a\in A}d(a,B),\, \sup_{b\in B}d(b,A) \}\,.
\end{equation}
\end{defn}

For non-compact manifolds one has to consider the more general {\it
Gromov-Hausdorff distance} (see, for example, \cite{BBI}).

Since, according to the above arguments, 
both the distance between $M$ and $\partial{\rm Tub}_\sigma^-(M)$,
$\partial{\rm Tub}_{\sigma}^+(M)$ and the deviations of their
curvature measures are arbitrarily small for small enough $\sigma$, we can state a first qualitative
geometric version of Shannon's Theorem for the Gaussian channel.
While a perfect analogy is not available, 
we can nevertheless formulate 
the following  theorem:

\begin{thm}
Let $M^n$ be a smooth geometric signal (manifold) and let $\sigma$
be small enough, such that  ${\rm Tub}_\sigma(M)$ is a submanifold
of $\mathbb{R}^{n+1}$. Then, given any noisy signal $M + Y$, such
that the average noise power $\sigma_Y$ is at most $\sigma$, there
exists a sampling scheme of $M + Y$ with an arbitrarily small probability
of resultant decoding error.
\end{thm}

The analogue of the capacity in the context of the geometric
approach to codes is $C_0 = C_0(n,\sigma,r)$, where $r$ represents
the differentiability class of $M$.

The existence of tube $\partial{\rm Tub}_\sigma^+(M)$ is, as noted,
guaranteed globally in the case of compact manifolds. Hence it
follows that the sampling scheme is also global and necessitates
$O(N)$ points, $N = N_M$. However, for non-compact manifolds (in
particular non-band limited geometric signals), the existence of
$\partial{\rm Tub}_\sigma^+(M)$ is guaranteed only locally.
Therefore ``gluing ''of patches is needed, an operation which
requires the insertion of additional vertices (i.e. sampling
points), their number being a function of the dimension of $M$.
Hence, in this case, $N_{M+Y} = O(N_M^n)$.

It is important to note that, again, this result is not restricted
to smooth manifolds, but rather extends to much more general
signals: Indeed, for any compact set $M \in \mathbb{R}^n$, the
$(n-1)$-dimensional sets
$\partial{\rm Tub}_\sigma^-(M)$, $\partial{\rm Tub}_\sigma^+(M)$, 
 are {\it Lipschitz manifolds}(i.e. topological manifolds equipped with a maximal atlas for which the changes of coordinates are Lipschitz functions), for almost any $\varepsilon$ (see \cite{HF}\,). Moreover,
the generalized curvatures measures of $\partial{\rm
Tub}_\sigma^-(M)$, $\partial{\rm Tub}_\sigma^+(M)$ are arbitrarily
close to the curvature of $M$, for small enough $\sigma$
(\cite{CMS1}, \cite{HF}). It follows that the above generalization
befits not only the case of the Gaussian noise, but to more general
types of noise, as well (see, \cite{Sh}, \cite{H}, \cite{LPM}).

Also, we note {\it en passant}, that practically the same argument as
above, with little (if any) modifications allows us to obtain a
geometric version of Landau's result on the
reconstruction of distorted signals, \cite{lan1}.

The full details of a {\it quantitative} version, including the
general case, are however, more involved and warrant a more detailed consideration elsewhere. 

Before we conclude this section, we wish to emphasize that the
importance of tubes is not necessarily limited 
to Differential Geometry. It is just as important in Statistics
(see, e.g. \cite{Lo}).
However, its relevance to sampling theory is in particular evident.
Indeed, Shannon's ideas, as exposed in \cite{Sh48} and \cite{Sh} are
very similar to our approach (even if lacking the specific geometric
nomenclature). In particular, convergence of the measure of
{entropy of the noise} introduced in \cite{Sh} is easily obtained in
the context of our approach and formalism. For instance, the equivalence of the tube radius is
used in \cite{Sh48} as a measure of the uncertainty of the
reconstruction (not to confused with the Heisenberg-related Uncertainty
Principle). For the development of Shannon's
approach see \cite{sz}.




\section{Appendix -- A Concise Proof of Theorem 1.1}

The proof of Theorem \ref{thm:main-smooth-hd} is essentially based on existence of fat triangulation for Riemannian manifolds.
In the sequel we outline some definitions and notations, and review relevant results concerning the existence of such triangulations.

Let $M^n$ denote an $n$-dimensional complete Riemannian manifold,
and let it be isometrically embedded into $\mathbb{R}^N$
(``$N$''-s existence is guaranteed by Nash's Theorem -- see, e.g.
\cite{Pe}).

Let $\mathbb{B}^\nu(x,r) = \{y \in \mathbb{R}^\nu\,|\, d_{eucl} < r\}$; $\partial\mathbb{B}^\nu(x,r) =
\mathbb{S}^{\nu-1}(x,r)$. If $x \in M^n$, let $\sigma^n(x,r) = M^n \cap \mathbb{B}^\nu(x,r)$, $\beta^n(x,r) =
exp_x\big(\mathbb{B}^n(0,r)\big)$, where: $exp_x$ denotes the exponential map: $exp_x:T_x(M^n) \rightarrow M^n$
and where $\mathbb{B}^n(0,r) \subset T_x\big(M^n\big)$, $\mathbb{B}^n(0,r) = \{y \in
\mathbb{R}^n\,|\,d_{eucl}(y,0) < r\}$.

The following definitions generalize in a straightforward manner classical ones used for surfaces in
$\mathbb{R}^3$:

\begin{defn}
\begin{enumerate}
\item $\mathbb{S}^{\nu-1}(x,r)$ is {\em tangent} to $M^n$ at $x\in M^n$ iff there exists $\mathbb{S}^n(x,r) \subset
\mathbb{S}^{\nu-1}(x,r)$, s.t. $T_x(\mathbb{S}^n(x,r)) \equiv T_x(M^n)$.
\item Let $l \subset \mathbb{R}^\nu$ be a line, then $l$ is {\em secant} to $X \subset M^n$ iff $|\,l \cap X| \geq 2$.
\end{enumerate}
\end{defn}

\begin{defn}
\begin{enumerate}
\item $\mathbb{S}^{\nu-1}(x,\rho)$ is an {\rm osculatory sphere} at $x \in M^n$ iff:
\begin{enumerate}
\item $\mathbb{S}^{\nu-1}(x,\rho)$ is tangent at x;
\\ and
\item $\mathbb{B}^n(x,\rho) \cap M^n = \emptyset$.
\end{enumerate}
\item Let $X \subset M^n$. The number $\omega = \omega_X = \sup\{\rho > 0\,|\, \mathbb{S}^{\nu-1}(x,\rho) \; {\rm osculatory} \\{\rm at\; any}\; x \in
X\}$ is called the {\em maximal osculatory} ({\em tubular}) {\em radius} at $X$.
\end{enumerate}
\end{defn}

\begin{rem}
There exists an osculatory sphere at any point of $M^n$ (see \cite{Ca}\,).
\end{rem}


\subsection{ Fat triangulations }

\begin{defn}\label{defn:FantTriang}
\begin{enumerate}
\item A triangle in $\mathbb{R}^2$ is called $\varphi$-{\em fat} iff  all its angles are larger than a prescribed value $\varphi> 0$.
\item A k-simplex $\tau \subset \mathbb{R}^n$, $2 \leq k \leq n$, is $\varphi${\em -fat} if there exists $\varphi > 0$ such that the ratio $\frac{r}{R} \geq \varphi$, where $r$ and $R$, are respectively, the radii of the inscribed and circumscribed (k-1)-spheres of $\tau$.
\item A triangulation $\mathcal{T} = \{ \sigma_i \}_{i\in \bf I }$ is {\it fat} if all its simplices are $\varphi$-{\em fat} for some $\varphi > 0$.
\end{enumerate}
\end{defn}

\begin{prop}[\cite{CMS}]
There exists a constant $c(k)$ that depends solely upon the
dimension $k$ of  $\tau$ such that
\begin{equation}   \label{eq:ang-cond}
\frac{1}{c(k)}\cdot \varphi(\tau) \leq \min_{\hspace{0.1cm}\sigma
< \tau}\measuredangle(\tau,\sigma) \leq c(k)\cdot \varphi(\tau)\,,
\end{equation}
and
\begin{equation}    \label{eq:area-cond}
\varphi(\tau) \leq \frac{Vol_j(\sigma)}{diam^{j}\,\sigma} \leq c(k)\cdot \varphi(\tau)\,,
\end{equation}
where $\varphi$ denotes the fatness of the simplex $\tau$, $\measuredangle(\tau,\sigma)$ denotes the  ({\em
internal}) {\em dihedral angle} of the face $\sigma < \tau$ and $Vol_{j}(\sigma)$; $diam\,\sigma$ stand for the
Euclidian $j$-volume and the diameter of $\sigma$, respectively. (If $dim\,\sigma = 0$, then $Vol_{j}(\sigma) = 1$,
by convention.)
\end{prop}

Condition  \ref{eq:ang-cond} is just the expression of fatness as
a function of dihedral angles in all dimensions, while Condition
\ref{eq:area-cond} expresses fatness as given by ``large
area/diameter''. Diameter is important since fatness is
independent of scale.

Existence of fat triangulations of Riemannian manifolds is guaranteed by the studies quoted below.

\subsubsection{Compact manifolds:} \label{s:Cairns}

In the seminal contribution of Cairns, \cite{Ca}, the following is proved,
\begin{thm}[\cite{Ca}] \label{thm:cairns}
Every compact $\mathcal{C}^2$ Riemannian manifold  admits a fat triangulation.
\end{thm}
\subsubsection{Open manifolds:}
\begin{thm}[\cite{Pe}] \label{s:Peltonen}
Every open (unbounded) $\mathcal{C}^\infty$ Riemannian manifold admits a fat triangulation.
\end{thm}
\subsubsection{Manifolds with boundary of low differentiability:}
\begin{thm}[\cite{Sa}] \label{s:Saucan}
Let $M^{n}$ be an $n$-dimensional $\mathcal{C}^{1}$ Riemannian manifold with boundary, having a finite number of
compact boundary components. Then, any  fat triangulation of $\partial M^{n}$ can be extended to a fat
triangulation of $M^{n}$.
\end{thm}

The method used to prove Theorem \ref{s:Cairns}, presented in \cite{Ca}, is to produce a point set $A \subseteq M^n$,
that is maximal with respect to the following density condition:

\begin{equation}
d(a_1,a_2) \geq \eta\,, {\rm for \; all\;}  a_1, a_2 \in A\,;
\end{equation}
where
\begin{equation}
\eta < \omega_M\,.
\end{equation}
One makes use of the fact that for a compact manifold $M^n$ we have $|A| < \aleph_0$, in order to construct the finite cell
complex ``cut out of M'' by the $\nu$-dimensional Dirichlet complex, whose (closed) cells $\bar{c}_k = \bar{c}_k^\nu$ are given by:
\begin{equation}
\bar{c}_k^\nu = \{x \in \mathbb{R}^\nu\,|\,d_{eucl}(a_k,x) \leq d_{eucl}(a_i,x),\;a_i \in A\,, \; a_i
\neq a_k\},
\end{equation}
i.e. the (closed) cell complex $\{\bar{\gamma}_k^n\}$, where:
\begin{equation}
\{\bar{\gamma}_k^n\} = \bar{\gamma}_k = \bar{c}_k \cap M^n \label{eq:gamma-n-k}
\end{equation}
For further details, see \cite{Ca}.

The proof of Theorem \ref{s:Peltonen}, \cite{Pe}, is based on adapting Cairns' method to the non-compact case.
Essential steps in the proof are as follows.

\begin{enumerate}[(i)]
\item Start with some compact submanifold $M_0^n$ of $M^n$.
\item Decompose $M^n$ as an exhaustion by open submanifolds, starting with $M_0^n$, namely present $M^n$ as,
\[ M^n = \bigcup_{j \in J}M_j^n \; ; \; M_j^n \subset M_{j+1}^n \; ;\; \overline{M_{j+1}^n} \setminus M_j^n \textrm{ is compact} \; . \]
\item Construct a fat triangulation of the initial compact submanifold $M_0^n$ according to \cite{Ca}, and iteratively extend it from $M_j^n$ to $M_{j+1}^n$ until a fat triangulation of the whole of $M^n$ is obtained.
\end{enumerate}

In order to prove Theorem \ref{s:Saucan} one first constructs two fat triangulations: $\mathcal{T}_{1}$ of a product
neighbourhood $N$ of $\partial M^n$ in $M^n$ and $\mathcal{T}_{2}$ of $int\, M^n$, the existence of which follows from
Peltonen's result \cite{Pe}, and then ``mashes'' the two triangulations into a new triangulation $\mathcal{T}$, while
retaining their fatness. While the mashing procedure of the two triangulations is basically the one developed in
the original proof of Munkres' theorem \cite{mun}, the triangulation of $\mathcal{T}_{1}$ has been modified, in order to ensure the fatness of the simplices of $\mathcal{T}_{1}$. The method we have employed for fattening triangulations
is the one developed in \cite{CMS}. For the technical details, see \cite{Sa}.

\subsection{From fat triangulations to sampling}
We first restate Theorem \ref{thm:main-smooth-hd}

{\it Theorem} (\cite{SAZ}):
{\it Let $\Sigma$ be a connected, non-necessarily compact smooth surface (i.e. of class $\mathcal{C}^k, k \geq 2$),
with finitely many boundary components. Then, there exists a sampling scheme of $\Sigma$, with a proper density
$\mathcal{D} = \mathcal{D}(p) = \mathcal{D}\!\left(\frac{1}{k(p)}\right)$, where $k(p) = \max\{|k_1|,|k_2|\}$, and
$k_1,k_2$ are the principal curvatures of $\Sigma$, at the point $p \in \Sigma$.
}

\begin{prf}
The sampling points are provided by the vertices of the fat
triangulation constructed above. The fact that the density is a function solely of $k = max\{|k_1|,...,|k_{n}|\}$
follows from theorem \ref{s:Cairns} \cite{Ca}, and from the fact that the osculatory radius $\omega_\gamma(p)$ at a
point $p$ of a curve $\gamma$ equals $1/k_\gamma(p)$, where $k_\gamma(p)$ is the curvature of $\gamma$ at $p$\,;
hence the maximal osculatory radius (of $\Sigma$) at $p$ is: $\omega(p) = \max\{|k_1|,...,|k_{n}|\} =
\max\{\frac{1}{\omega_1},...,\frac{1}{\omega_{n}}\}$. (Here $\omega_{i}\,, i = 1,...,n$ denote the minimal,
respective maximal sectional osculatory radii at $p$.)
\end{prf}

Since for unbounded surfaces it may well be that $\kappa \rightarrow \infty$, it follows that an infinite density
of the sampling is possible. However, for practical implementations, where such cases are excluded, we have the
following corollary:

We also quote here the following immediate corollary derived from Theorem \ref{thm:main-smooth-hd}
\begin{cor} [\cite{SAZ}]\label{cor:cor1sampling}
Let  $\Sigma, \mathcal{D}$ be as above. Assume that there exists $k_0 > 0$, such that $k_0 \geq k(p)$, for all $p
\in \Sigma$. Then, there exists a sampling of $\Sigma$ having uniformly bounded density.
\end{cor}

In the following cases there exist $k_0$ as in the above Corollary \ref{cor:cor1sampling} (\cite{SAZ}):
\begin{enumerate}
\item $\Sigma$ is compact.
\item There exist $H_1,H_2,K_1,K_2$, such that $H_1 \leq H(p) \leq H_2$ and $K_1 \leq K(p) \leq K_2$, for any  $p \in \Sigma$, where
$H,K$ denote the mean, respective Gauss curvature. (That is both mean and Gauss curvatures are {\em pinched}.)
\item The {\it Willmore integrand} $W(p) = H^2(p) - K(p)$ and $K$ (or $H$) are pinched.
\end{enumerate}

Theorem \ref{thm:main-smooth-hd} is valid for non-smooth surfaces, (\cite{SAZ}).

\subsection{Reconstruction}
\begin{defn}[\cite{mun}]
\begin{enumerate}
\item Let $f:K \rightarrow \mathbb{R}^n$ be a $\mathcal{C}^r$ map, and let $\delta:K \rightarrow \mathbb{R}^*_+$ be a continuous function. Then $g:|K| \rightarrow \mathbb{R}^n$ is called a $\delta${\em-approximation to} $f$ iff:\\
(i) There exists a subdivision $K'$ of $K$ such that $g \in
\mathcal{C}^r(K',\mathbb{R}^n)$\,;\\
(ii) $d_{eucl}\big(f(x),g(x)\big) < \delta(x)$\,, for any $x \in |K|$\,;\\
(iii) $d_{eucl}\big(df_a(x),dg_a(x)\big) \leq \delta(a)\cdot
d_{eucl}(x,a)$\,, for any $a \in |K|$ and for all $x \in \overline{St}(a,K')$.
\item Let $K'$ be a subdivision of $K$, $U = \raisebox{0.05cm}{\mbox{$\stackrel{\circ}{U}$}}$, and let $f \in
\mathcal{C}^r(K,\mathbb{R}^n), \; g  \in \mathcal{C}^r(K',\mathbb{R}^n)$.  g is called a
$\delta${\em-approximation} of $f$ (on $U$) iff conditions (ii) and (iii) above hold for any $a \in
U$.
\end{enumerate}
\end{defn}
\begin{defn}[\cite{mun}] \label{s:Secant}
Let $f \in \mathcal{C}^r(K)$ and let $s$ be a simplex, $s < \sigma \in K$. Then the linear map: $L_s:s \rightarrow
\mathbb{R}^n$, defined by $L_s(v) = f(v)$ where $v$ is a vertex of $s$, is called the {\em secant map induced by}
$f$.
\end{defn}

The motivation for having fat triangulations for manifolds in terms of reconstruction of sampled manifolds is stressed by the following theorem.

\begin{thm} [\cite{mun}]\label{s:Munkres}
Let $f:\sigma \rightarrow \mathbb{R}^n$ be of class $\mathcal{C}^k$. Then, for $\delta > 0$, there exists
$\varepsilon, \varphi_0 >0$, such that, for any $\tau < \sigma$, fulfilling the following conditions:\\
(i) $diam(\tau)\; < \; \varepsilon$ and,\\
(ii) $\tau$ is $\varphi_0$-fat,\\
then, the secant map $L_\tau$ is a $\delta$-approximation of $f|\tau$.
\end{thm}

Following Theorem \ref{s:Munkres}, we use the secant map as defined in Definition \ref{s:Secant}, in order to reproduce a PL-manifold as a $\delta$-approximation for the sampled manifold. We may
now use smoothing in order to obtain a $\mathcal{C}^{\infty}$ approximation.

It should be emphasizes here that Theorem \ref{thm:main-smooth-hd} produces an infinite sequence of $PL$-approximations of the sampled manifold which, as already stated above, converge, when the diameter of simplices tends to zero, to the original sampled manifolds. Moreover, following \cite{CMS}, we also have that the discrete curvature measures given on the approximating manifolds converge, as measures, to the sectional curvature of the sampled manifold. This is in accordance with the fact that, in order to achieve proper reproducing of a $1$-dimensional signal according to Shannon Theorem, as in \cite{Sh}, one has to account for an infinite summation.


\subsection*{Acknowledgment}
The first author would like to thank Professors Shahar Mendelson and
Gershon Wolansky -- their warm support is gratefully acknowledged.


The third author wishes to thank Professor Shlomo Shamai for pointing out
that the geometric sampling approach actually motivated Shannon, and
that this approach is inherent  in his pioneering work.




\end{document}